\documentclass[elsart]{elsart}
\usepackage{graphicx,amssymb,amsmath,times}
\journal{Astroparticle Physics}
\begin{document}
\begin{frontmatter}
\title{Cosmic-ray composition with TACTIC telescope
 using Fractal and Wavelet Analysis}
\author{C.K.Bhat}
\address{Astrophysical Sciences Division, Bhabha Atomic Research Centre\\ Mumbai-400085, India}
\ead{ckbhat@barc.gov.in}

\begin{abstract}  A preliminary flux estimate of various cosmic-ray constituents 
based on the atmospheric Cerenkov light flux of extensive air showers using fractal and wavelet analysis approach is proposed. Using a Monte-Carlo simulated database of Cerenkov images recorded by the TACTIC telescope, we show that one of the wavelet parameters (wavelet dimension $B_{6}$) provides $\geq$ 90\%\ segregation of the simulated events in terms of the primary mass. We use these results to get a preliminary estimate of primary flux for various cosmic-ray primaries above 5 TeV energy. The simulation based flux estimates of the primary mass as recorded by the TACTIC telescope are in good agreement with the experimentally determined values.
\end{abstract}
\end{frontmatter}

\section{Introduction}
One of the main problems in the field of primary cosmic-ray investigations is the form of the energy spectrum and the precise determination of the mass composition. Precise information about the energy spectrum and mass composition of primary cosmic-rays at VHE ($\geq$ 10$^{11}$ eV) energies is an important input for understanding the origin of cosmic-rays and their propagation through the Galaxy. Cosmic-rays exhibit a nearly smooth power law energy spectrum over more than 10 decades of energy (from $\sim$ 10$^{9}$ to 10$^{20}$eV) with significant kinks at $\sim$ 10$^{15}$ eV (knee) and $\sim$ 10$^{18}$eV (ankle)[1,2]. The knee feature at $\sim$ 10$^{15}$eV, which involves a change in the differential spectral exponent from  $\sim$ 2.7 to $\sim$ 3.1 has defied a clear explanation so far. Among the possibilities, one of the reasons suggested for the knee feature involves a change in the composition of the primary cosmic-rays from a mainly proton dominated($\sim$ 90\%\ protons) to a mixed (protons + heavy nuclei) composition or even heavy nuclei domination[2,3]. Thus, it becomes very important to measure the cosmic-ray composition in the VHE range, especially at energies around 10$^{14}$-10$^{15}$eV. Although primary cosmic-ray composition is known more or less precisely at energies up to $\sim$ 10 TeV from direct satellite/balloon based observations[4,5], the low flux of primary cosmic-ray particles at energies $\geq$ 10$^{13}$ eV ($\sim$ 10$^{-2}$particles cm$^{-2}$ s$^{-1}$)does not allow direct measurement of primary composition with detectors carried on balloons and satellites. Attempts have, therefore, been made to infer the cosmic-ray composition indirectly from data collected with large air shower arrays and atmospheric Cerenkov telescopes which detect primary cosmic-rays indirectly by recording the secondary particles and photons produced in the atmospheric cascades initiated by these primaries[6,7,8,9]. These attempts have achieved limited success so far because of the difficulty in relating the observed cascade parameters to the primary mass. In the present communication, we make an attempt to show that one of the fractal/wavelet parameters derived from the Cerenkov images recorded by an imaging Cerenkov telescope (like TACTIC) exhibits an unprecedented efficiency for separating primary cosmic-rays in terms of their mass. This makes it possible to estimate the flux of cosmic-ray primaries at energies $\geq$ 10$^{13}$eV at large zenith angles with ground-based atmospheric Cerenkov telescope systems.

\section{Imaging Atmospheric Cerenkov Technique}  Primary cosmic-rays (protons, hadrons, photons etc.) with energy $\geq 10^{11}$eV, initiate extensive air showers in the atmosphere, generating in the process a large number of secondary particles ($\e^{\pm},\mu^{\pm},\pi^{\pm}$, hadrons, photons) some of which survive absorption in the atmosphere and can be detected at ground level. A large number of $\e^{\pm}$, traveling with relativistic speeds, generate Cerenkov photons in the atmosphere which arrive in the form of a few ns duration optical pulse ($\lambda$$\sim 300- 600 nm$) spread over a radius of $\sim$ 100m at ground level[6,7,10]. An imaging Cerenkov telescope, comprising a large light collector with a matrix of photomultiplier tube (PMT) detectors in its focal plane, generates a two-dimensional picture of the resultant Cerenkov photon distribution (called the Cerenkov image ), the geometrical features of which depend upon the nature (mass) of the cosmic-ray primary which initiated the cascade. The technique has been successfully employed by various imaging Cerenkov telescopes including TACTIC to detect VHE gamma-ray signals from a number of galactic and extragalactic sources[7,8]. The TACTIC (TeV Atmospheric Cerenkov Telescope with Imaging Camera) telescope at Mt.Abu (24.6$^{\circ}$ N; 72.0$^{\circ}$ E; 1300m asl) comprises a 3.6m aperture, tessellated light collector of overall parabolic shape mounted on an altitude-azimuth mount and equipped with a 349-pixel imaging camera at its focal plane [10,11]. The imaging camera provides a field of view of 3$^{\circ}$ radius with  a uniform pixel resolution of 0.31$^{\circ}$. Using a topological trigger based on a coincidence($\tau$ $\sim$ 20ns) between nearest-neighbour, non-collinear triplets (3NCT) in the central 225 pixels (15{$\times$}15), the imaging telescope records Cerenkov images of extensive air showers initiated by primary gamma-rays of $\geq$ 1.5 TeV and primary hadrons of energy $\geq$ 3 TeV from the vertical direction[10]. The images are composed of Cerenkov photon intensity 
observed by each pixel in the form of charge to digital converter counts(CDC). The effective collection area (area of Cerenkov light pool at ground level) of the system has been estimated to be $\sim$ 3${\times}$10 $^{8}$ cm$^{2}$[17]. Both the threshold detection energy and the effective detection area increase with the zenith angle of observation[10,12].

\section {Simulations} 
The results presented in this paper are based on the analysis of simulated Cerenkov images produced by several primary species like gamma-rays, protons, Ne and Fe nuclei. The simulations have been performed for the 349$-$pixel imaging camera of the TACTIC telescope using the CORSIKA (version 6.51) air shower simulation code[14,15,16], supplemented by appropriate custom built software to take into account atmospheric absorption, reflection at the TACTIC light collector surface, PMT characteristics and the night sky background induced shot-noise fluctuations[7,13]. For each primary species 6,00,000 showers in different energy ranges ( 2TeV, 4TeV, 5TeV, 8TeV, 10TeV, 15TeV, 20TeV, 30TeV) with energy spectral index of -2.5, -2.76, -2.64 and  -2.60 were simulated for gamma-rays, protons, Neon and Fe respectively. All primary particles were assumed to be incident isotropically within the $\sim$ 4$^{\circ}$diameter field of view of the trigger pixels along a zenith direction of 50$^{\circ}$. 

The large value of zenith angle was deliberately chosen to obtain a larger effective area of detection with a corresponding increase in the energy threshold as mentioned earlier. Independent simulations of the TACTIC telescope carried out earlier have shown that the detection threshold for gamma-rays at $\geq$ 45$^{\circ}$ zenith angle increases to $\sim$ 10 TeV (2 TeV for vertical) while the effective collection area increases to $\sim$ 10$^{9}$ cm$^{2}$[12,17]. In this simulation study, we  have also evaluated the effective collection area for 5TeV gamma-rays for the TACTIC telescope at zenith direction of 50$^{\circ}$ as $\sim$ 7.07$\times$ 10$^{8}$ cm$^{2}$. All the simulated Cerenkov images were subjected to similar cleaning and parameterization procedure in order to derive the Hillas image parameters [18] and the recently introduced fractal and wavelet parameters as described briefly below.

\subsection {Image Shape(Hillas)Parameterization} 
The simulated Cerenkov images recorded at core distances varying from 10-400m have been parameterized following the well-known second order moment analysis of the derived photo-electron counts in the image pixels. Hillas shape parameters[18], namely Length (L), Width(W), Azwidth (A), Distance (D), Miss (M), Orientation angle($\alpha$) and size (S) were determined for each simulated image. The derived values were sorted as an independent set, for core distances $\leq$ 400m for a comparative analysis.

\subsection{Fractal and Wavelet Parameters} 
The longitudinal and lateral development of air showers, represented by the height of shower maximum, the number of relativistic electrons and muons, the angular and spatial distribution of Cerenkov photons and their total number varies with the primary mass. This produces differences in the structure of Cerenkov images which can be recognized through pattern recognition techniques. Fractal and wavelet parameters quantify the Cerenkov image structure on different scale lengths, the former characterizing the global structure of photon distribution in the image while the latter characterize more localized structures[19,20,24,25]. The basic motivation and the technique employed for fractal and wavelet analysis of Cerenkov images has been described in detail in[21-25]. We have followed an essentially similar procedure to calculate fractal moments and dimensions as well as wavelet moments and dimensions of different order.

\subsection{Fractal and Wavelet Analysis of Cerenkov Images}
There has been increasing interest in wavelet based approach in the recent years due to its success in several applications. The wavelet transform is a two parameter transform[26-28] . The two parameter domains of the wavelet tranform are dilation (scaling or compression) and translation (space). The advantage of using wavelets in the signal or structure analysis is the ability to perform local analysis of the signal or structure i.e to zoom on any interval of time and space. As such wavelet analysis has the ability to reveal the subtle details of the data which other analysis techniques fail to detect. Wavelet Analysis is regarded as the sequence of versatile filtering approach which will allow to examine the very minute stuctures left after filtering of the images at different scale lengths i.e to zoom at any interval of frequency ,time or space of the signal and enables us to detect some hidden aspect of the data or image and brings out corrections to the image analysis. 

The pattern recognition technique for gamma-hadron separation employs fractal and wavelet analysis of Cerenkov images to exploit differences in the structure of the recorded images due to differences in the longitudinal and lateral development of showers initiated by the two primary species. Showers initiated by primary gamma-rays and protons produce different numbers of relativistic electrons and muons and also have different heights (and lateral extents) of shower maximum. These differences lead to structures in the Cerenkov image on different scale lengths which can be parameterized through multifractal and wavelet dimensions and moments  of different order[19-24] . The usual procedure for calculating these parameters involves dividing the Cerenkov image into M = 4, 16, 64, 256 equally sized, non-overlapping cells and finding the number of CDC counts in each cell (6.5 CDC = 1 photo-electron). The multifractal moments of order q(q = 2$-$ 6) are calculated from the following relation: 

\begin{eqnarray}
\textrm{Fractal moment} &G_{q}(M) &= \sum_{j=1}^{M}\left[\frac{K_{j}}{N}\right]^{q} 
\end{eqnarray}

where N is the total number of CDC counts in the image, $K_{j}$ is the CDC count in the j$^{th}$ cell and q= 2, 3, 4, 5, 6 is the order of fractal moment. The wavelet moments are obtained as follows:
\begin{eqnarray}
\textrm{Wavelet moment} &W_{q}(M)&=\sum_{j=1}^{M}\left[(\frac{K_{j+1}-K_{j}}{N})\right]^{q} 
\end{eqnarray}

The fractal scale length is given by ${\nu}$ = $log_{2}$M 

Both the fractal moment $G_{q}$ as well as the wavelet moment $W_{q}$ exhibit a proportionality with the scale length which can be expressed as,

\begin{equation}
  G_{q}(M) \propto M ^{\tau_q}\qquad\textrm{and}\qquad\ 
  W_{q}(M) \propto M ^{\beta_q} 
\end{equation}
    
The fractal and wavelet dimensions of order q are then determined from the following relations:
 
\begin{eqnarray}
\textrm{Fractal dimension}   &D_{q} &= \frac{\tau_q}{q-1} \\
\textrm{Wavelet dimension}   &B_{q}&= \frac{\beta_q}{q-1}
\end{eqnarray}

where $\tau_{q}$ and $\beta_{q}$ are the slopes of the line between natural logarithm with scale length $\nu$ of $G_{q}$(M) and $W_{q}$(M) respectively. Since wavelet dimensions are more sensitive to local structures in the photon distribution in Cerenkov images, they are expected to have better efficiency for gamma-hadron separation as proton-initiated showers are known to have a non-uniform structure due to contributions from pions and muons[20-25]. 

\subsection{Results of Image Parameterization and Fractal-Wavelet Analysis}
Fig.1 shows a composite plot of the frequency distribution of image shape parameters derived for showers initiated by 5-20 TeV gamma-rays, protons, Ne and Fe primaries with shower cores located at $\leq$ 400m from the TACTIC imaging telescope. We find that the distributions for the shape parameters length and width (L,W) show a clear distinction between gamma-ray, proton and Ne/Fe induced showers, especially a clear distinction between gamma-ray and hadron initiated showers. While on the other hand, the distributions for the Alpha($\alpha$), Distance(D) and Miss(M) parameters do not exhibit any substantial dependence on the primary mass indicating that these parameters can have a limited application in primary mass segregation.

\begin{table}        
\centering 

\caption {Parameter domains and retention factors of different primary species for Cerenkov events at $\leq$ 400m; A-Accepted, R-Rejected}

\begin{tabular}{|p{1.0cm}|c|c|c|c|c|}
\hline
 -&-&\%\      & \%\      & \%\       &\%\       \\

 S No &  Domain  & $\gamma$ &  p &  Ne &  Fe\\
\hline        
 1  &  0.06$\leq$ L$\leq$ 0.46   &   91.0A  &   95.0R  &  97.0R  &  99.0R\\ 
    &  0.06$\leq$ W$\leq$ 0.26 & & & &\\
\hline
2  &  0.30$\leq$ L$\leq$ 0.58    &  96.0R  &  87.0A  &  96.0R  &  95.0R\\ 
     &  0.20$\leq$ W$\leq$ 0.48  & & & &\\ 
\hline   
 3   & 0.40$\leq$ L$\leq$ 0.89    &  99.5R  &  80.0R  &  48.0A  &  78.5R\\  
     & 0.40$\leq$ W$\leq$ 0.62 & & & &\\
\hline
  4   &  0.54$\leq$ L$\leq$0.87   & 99.5R  &  82.5R  &  66.0R &  37.0A\\ 
    &  0.45$\leq$ W$\leq$0.69    & & & &\\
\hline   
\end{tabular}
\end{table}

\begin{figure}[h]
\centering
\includegraphics*[width=1.0\textwidth,angle=0,clip]{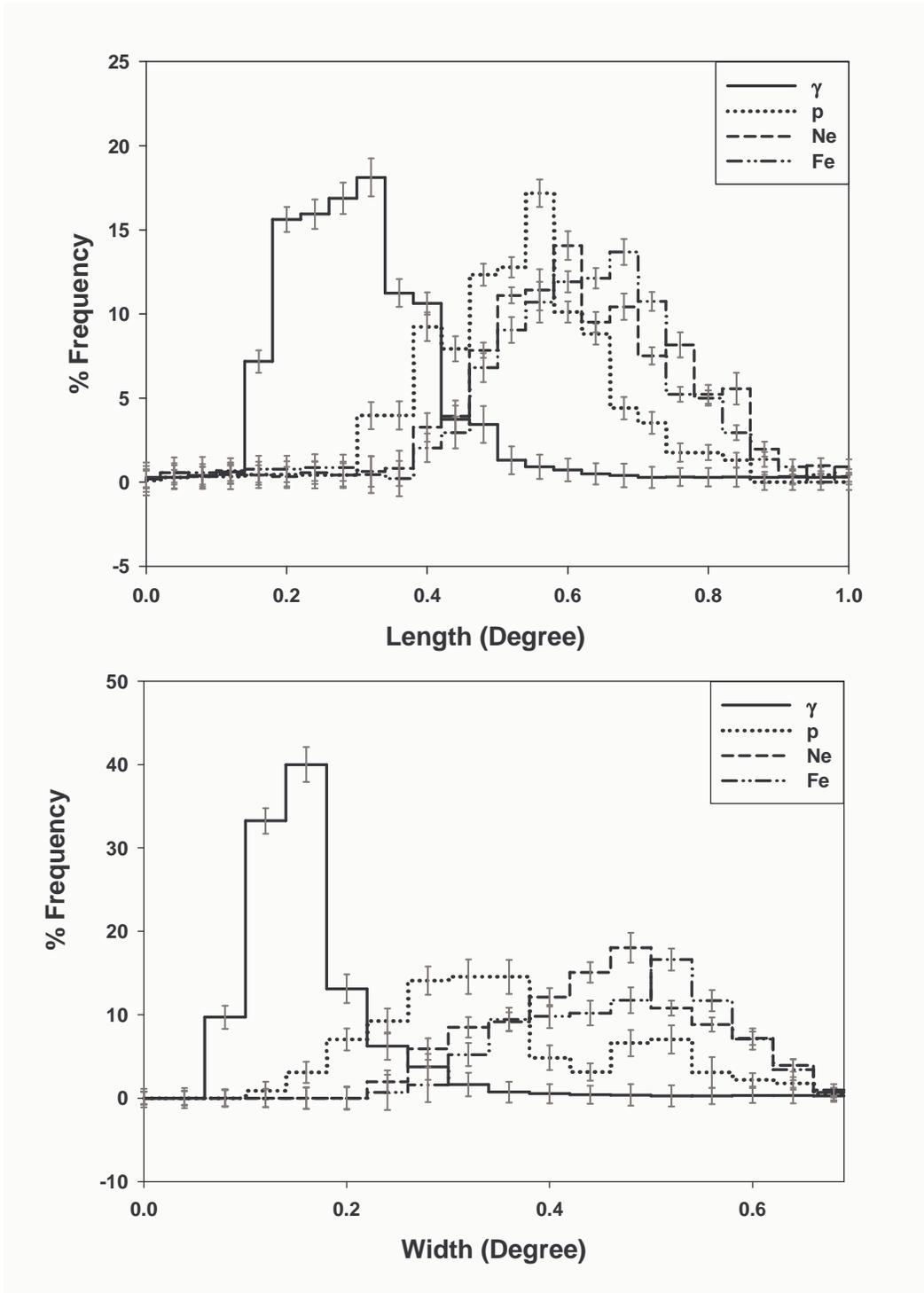}

\caption{\label{fig1}Image parameter distributions for gamma and hadron-initiated showers}
\end{figure}

This is expected because gamma-ray initiated Cerenkov images are known to be compact in shape as compared to the rather diffuse images produced by hadron initiated showers due to the presence of a large number of muons in the latter. For core distances $\leq$ 400m, only Length(L), Width(W) parameters show the above mentioned difference while the other distributions tend to show no segregation. In Hillas parameterization as depicted in Table.1, L and W parameter cuts show segregation between gammas and hadrons (p, Ne and Fe). Applying these cuts with appropriate values of L and W at different levels on Cerenkov images recorded by the telescope yields an acceptance for gamma-rays $\sim$ 91\%\ , for p $\sim$ 87\%\ , for Ne $\sim$ 48\%\  and for Fe $\sim$ 37\%\ .  We have tried various data cuts in order to find the optimum parameter domain which preferentially selects the maximum number of events initiated by one primary species alone. The results summarized in Table.1 show that, on the basis of data cuts involving L, W parameters, it is possible to broadly segregate the simulated events in terms of the primary mass. While this may not be useful  as  far as  determining the absolute cosmic-ray mass composition is concerned, this procedure can be used to at least know if the cosmic-rays undergo enrichment by heavy nuclei as the energy increases beyond $\sim$ 10 TeV.\\    

As mentioned earlier the simulated Cerenkov images produced by the four primary species $\gamma$, p, Ne, Fe of different energies  have also been subjected to analysis to calculate fractal and wavelet moments $G_{q}$ and $W_{q}$ as well as the multifractal and wavelet dimensions $D_{q}$  and $B_{q}$  of order q(-6 $\leq$q$\leq$ 6)[17-20]. The peak value of fractal dimension $D_{6}$ for the simulated Cerenkov images of $\gamma$, p, Ne, Fe lies at (0.4$\pm$0.02), (0.6$\pm$0.02), (0.7$\pm$0.03) and (0.8$\pm$0.03) respectively. Similarly $B_{6}$ values for these images lie at (5$\pm$0.6) for $\gamma$, (2.0$\pm$0.8) for p, (3.0$\pm$0.6) for Ne and (4$\pm$0.6) for Fe. The frequency distributions of $D_{q}$ and $B_{q}$ have been plotted in order to find any differences which could be exploited to segregate the events in terms of the primary mass. In this paper, we have considered the entire range of cosmic primaries i.e diffuse gamma-rays, protons, Ne and Fe. The gamma-rays considered in our simulation are within a field of view of $\sim$ 4$^{\circ}$.   Fig.2 and Fig.3 show a composite representative frequency distribution of fractal and wavelet parameters namely, $D_{2}$, $D_{6}$, $B_{2}$ and $B_{6}$. A careful visual examination of Fig.2 reveals that while the distributions for fractal parameters $D_{2}$ and $D_{6}$ exhibit a clear difference between events of $\gamma$-ray and Fe origin with events initiated by p and Ne primaries falling in between, they do not promise high efficiency for segregating the events on the basis of their mass as the distributions are rather wide with significant overlaps among the four distributions due to large background contribution. 

It may be mentioned here that the same general features are exhibited by the distributions for fractal dimensions of different order q, indicating that the fractal parameters have poor efficiency for mass segregation of the Cerenkov events. On the contrary, the frequency distributions for the wavelet parameters are found to be very narrow and peaked. In one particular case they are distinctly separated from one another for the four primary species considered here. This is seen clearly in Fig.3 where we have shown two representative distributions of wavelet dimensions $B_{2}$ and $B_{6}$. We find that the frequency distributions of $B_{6}$ for 5 -20 TeV $\gamma$, p, Ne and Fe initiated events are completely segregated from one another with $\leq$ 1\%\ gamma- events having $B_{6}$ =(5$\pm$0.6), $\leq$ 1\%\ Ne events having $B_{6}$ =(3.0$\pm$0.6), $\geq$ 87\%\ p events having $B_{6}$ =(2.0$\pm$0.8) and $\leq$ 1\%\ Fe events with $B_{6}$ =(4.0$\pm$0.6). The distributions for wavelet dimensions of other orders exhibit the same broad features as seen in the frequency distribution of $B_{2}$ indicating that, while these parameters can be used for $\gamma$- p segregation, they can not be employed for assigning a particular primary mass to the events due to large overlapping of background events. As is clear from Table.2, We find that the wavelet dimension $B_{6}$ provides the maximum efficiency for segregating the Cerenkov events in terms of their mass on an event by event basis, making it possible to study cosmic-ray composition in the multi-TeV energy range with an imaging atmospheric Cerenkov telescope.

The results so far obtained from simulations in each domain of $B_{6}$ initiated by these four cosmic-ray primaries were used to estimate their flux values at different energies. Various domains of wavelet dimensions $B_{2}$ and $B_{6}$ for each species have been identified from the simulated data. Out of 2,76,904 simulated events, 2,40,903$\pm$491 events were retained in the proton domain with $B_{6}$ = (2.0$\pm$0.8); 2,492 $\pm$ 44 events are falling in Ne domain with $B_{6}$ =(3.0$\pm$0.6); 2,216 $\pm$ 41 events are recorded in Fe domain with $B_{6}$ = (4$\pm$0.6) and 155 $\pm$ 11 events are recorded in the $\gamma$ domain with $B_{6}$ = (5$\pm$0.6). These values lead to the flux estimation of (3.8$\pm$0.5)$\times$ 10$^{-3}$ m$^{-2}$ s$^{-1}$ sr$^{-1}$, (4.3$\pm$0.5)$\times$ 10$^{-5}$ m$^{-2}$ s$^{-1}$ sr$^{-1}$, (3.9$\pm$0.5)$\times$10$^{-5}$ m$^{-2}$ s$^{-1}$sr$^{-1}$ and (2.9$\pm$0.4)$\times$10$^{-6}$ m$^{-2}$ s$^{-1}$ sr$^{-1}$ for protons, Ne, Fe and gamma-rays above 10TeV, 15TeV, 20TeV, and 5TeV energy respectively..

\begin{figure}[h]
\centering
\includegraphics*[width=1.0\textwidth,angle=0,clip]{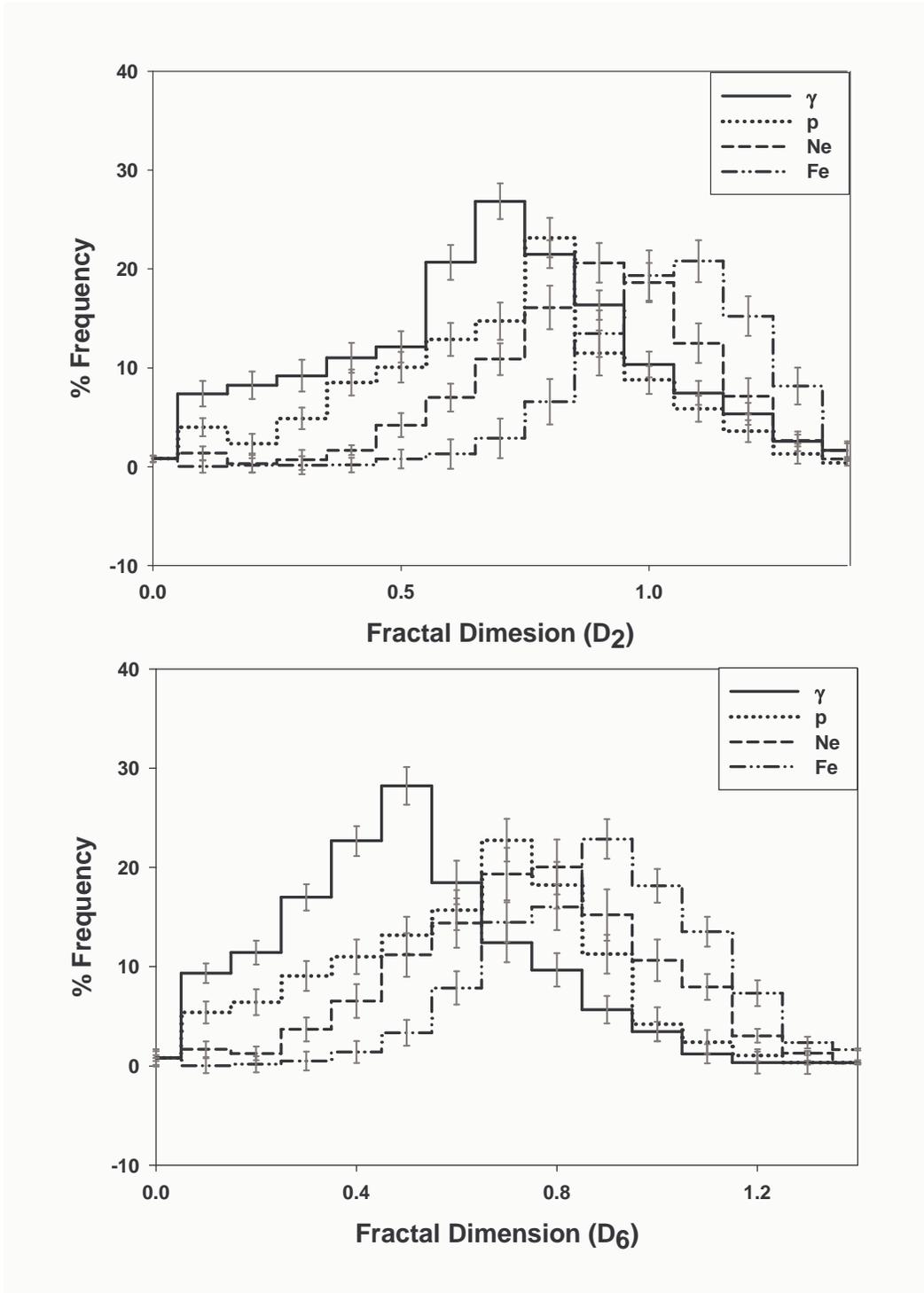}
\caption{\label{fig2}Fractal dimension distributions of Cerenkov
images generated by gammas-rays and hadrons (p,Ne,Fe)} 
\end{figure}

\begin{table}        
\centering 
\caption {Parameter domains and retention factors of different primary species for events at various $B_{6}$ values; A-Accepted, R-Rejected}

\begin{tabular}{|p{1.0cm}|c|c|c|c|c|}

\hline

 -&-&\%\      & \%\      & \%\       &\%\       \\

 S No &  Domain  & $\gamma$ &  p &  Ne &  Fe\\

\hline

 1  &  1.6$\leq$ $B_{6}$ $\leq$ 2.4   &   99.0R  &   98.5A  &  99.0R  &  99.5R\\ 
 
\hline

2  &  2.8$\leq$ $B_{6}$ $\leq$ 3.6    &  99.0R  &  98.5R  &  96.0A  &  98.5R\\
     
\hline

 3   & 3.8$\leq$ $B_{6}$ $\leq$ 4.6    &  98.5R  &  99.0R  &  97.5R  &  95.0A\\  
    
\hline

  4   &  5.2$\leq$  $B_{6}$  $\leq$5.6   & 91.0A  &  98.0R  &  97.0R &  96.0R\\ 
\hline   
\end{tabular}
\end{table}

\section{Experimental Results} 
The Cerenkov imaging telescope TACTIC has been used to search for TeV gamma-ray signals from several galactic and extragalactic sources and has successfully detected emission from the Crab Nebula, Mkn 421 and Mkn 501[29,30,31]. The telescope is also used to monitor an off-source or background region where it tracks a region of sky away from the source direction in order to provide an independent measure of the cosmic-ray background rate and also to validate the gamma-ray signal retrieval methodology. The fractal dimensions obtained from these Cerenkov images showed no potential in terms of segregation of events on the basis of mass because of significant overlaps of distributions due to large background contribution. These simulated event values also yield a cosmic-ray proton composition of 87\%\ which is close to the experimentally measured proton abundance of $\sim$ 89\%\ at TeV energies. The gamma-ray to proton ratio determined from these values is also in close agreement with the value of 0.001 at TeV energies reported from other atmospheric Cerenkov telescope systems[32] while as the protons out number other species by few orders of magnitude. 

We have also subjected Cerenkov images of various cosmic-ray primaries recorded by TACTIC during  off-source observation runs to the above discussed fractal and wavelet analysis to investigate whether the $B_{6}$ parameter distribution leads to the expected cosmic-ray mass composition. Fig.4 shows the frequency distribution of parameter $B_{6}$ for both simulated and experimentally observed Cerenkov images recorded by the TACTIC telescope during a 37h long off-source exposure along a direction making a zenith angle of $\sim$ 50$^{\circ}$. The total number of events recorded during Crab-off source scans was 2,76,900 corresponding to the event rate of $\sim$ 2.0 Hz, which translates into detection threshold energy of 4 TeV for cosmic-ray proton initiated events($\sim$ 2 TeV for gamma-ray initiated events) as mentioned in the simulation process. While 2,35,635$\pm$485 events fall within the proton domain $B_{6}$(2.0$\pm$0.8); 1,788 $\pm$ 39  events are falling in Ne domain with $B_{6}$= (3.0$\pm$0.6); 1,538 $\pm$ 34  events are recorded in Fe domain with  $B_{6}$ = (4$\pm$0.6)  while as 127 $\pm$ 9 events fall in the gamma domain with $B_{6}$=(5$\pm$0.6). Taking into account the exposure time (37h), TACTIC effective area  7.07$\times$ 10$^{8}$ cm$^{2}$ and the TACTIC field of view 3.8$\times$ 10$^{-3}$sr, the flux for various cosmic-primaries such as p, Ne, Fe and gammas above 5 TeV within 40\%\ systematic error are found to be (3.7$\pm$0.5)$\times$ 10$^{-3}$m$^{-2}$s$^{-1}$ sr$^{-1}$, (3.0$\pm$0.5)$\times$ 10$^{-5}$m$^{-2}$ s$^{-1}$ sr$^{-1}$, (2.7$\pm$0.4)$\times$ 10$^{-5}$m$^{-2}$ s$^{-1}$sr$^{-1}$ and (2.4$\pm$0.4)$\times$10$^{-6}$m$^{-2}$ s$^{-1}$ sr$^{-1}$ respectively[29]. The results obtained show good agreement between the simulated and experimental data. 

\begin{figure}[h]
\centering
\includegraphics*[width=1.0\textwidth,angle=0,clip]{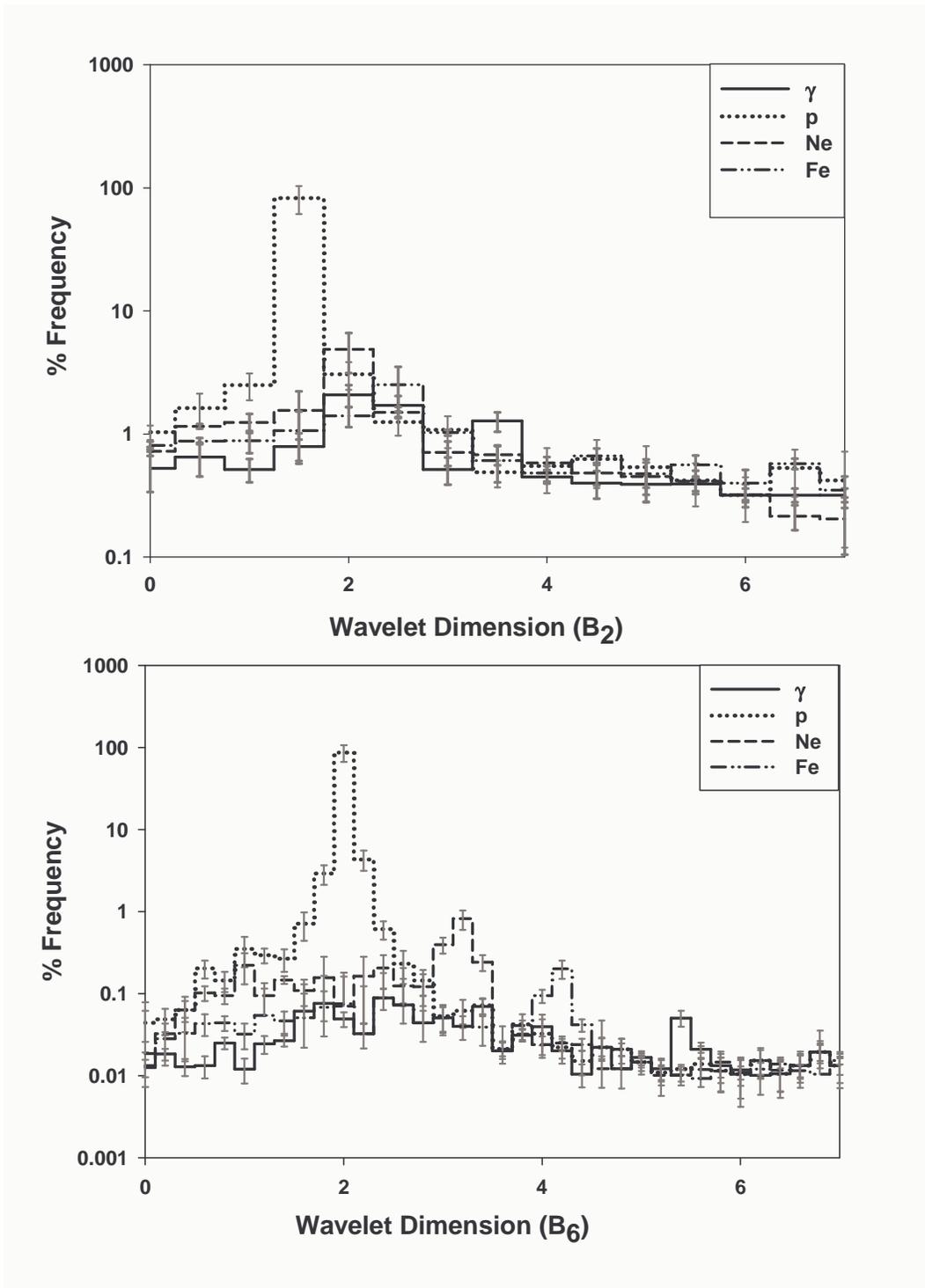}
\caption{\label{fig3}Wavelet dimension distributions of Cerenkov
images generated by gammas-rays and hadrons (p,Ne,Fe)} 
\end{figure}

\begin{figure}[h]
\centering
\includegraphics*[width=1.0\textwidth,angle=0,clip]{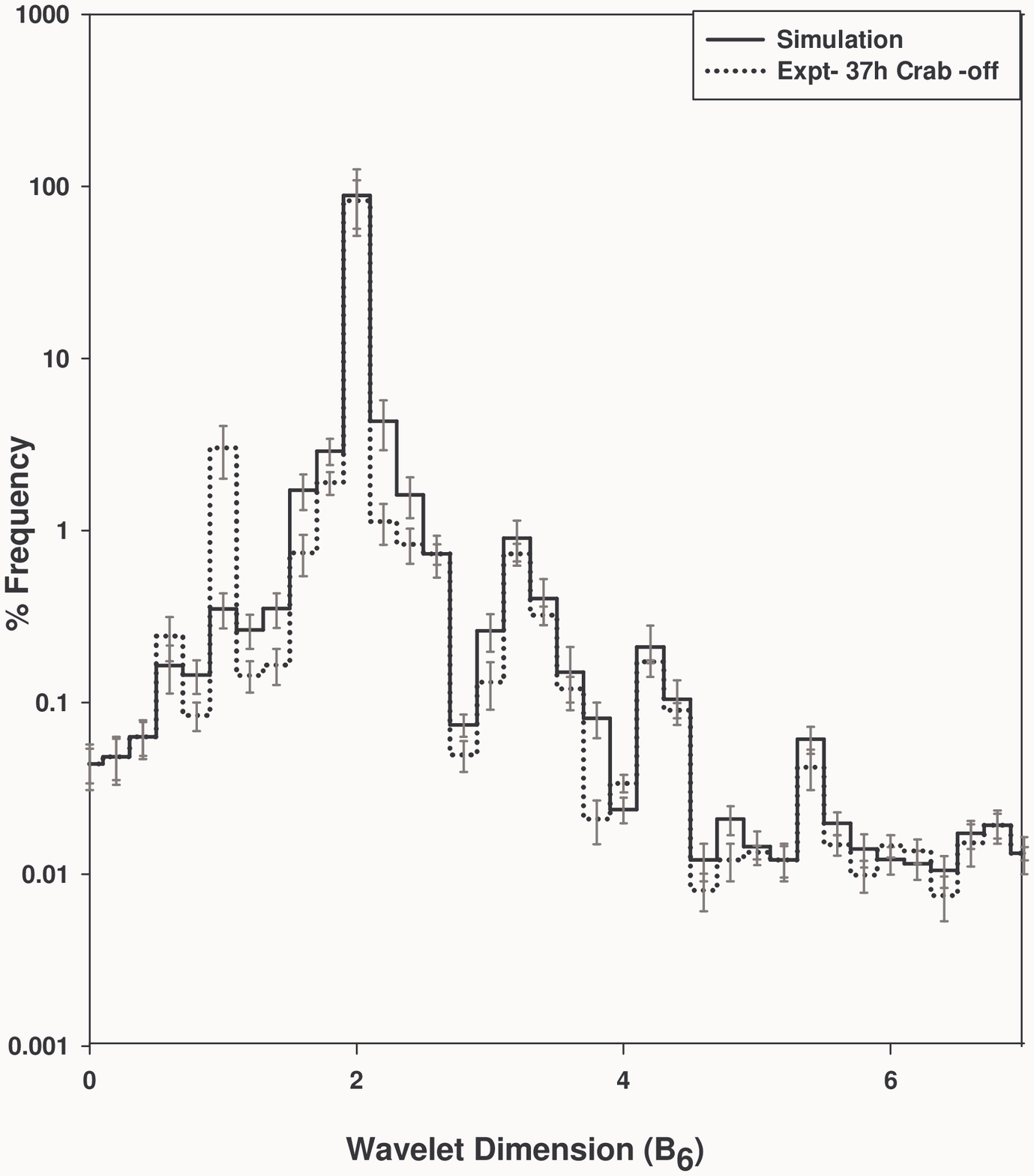}
\caption{\label{fig 4}Wavelet dimension distribution of simulated and experimental data} 
\end{figure}
                 
\section{Summary and Conclusions} 
 A number of attempts have been made earlier to assess the potential of imaging atmospheric Cerenkov telescopes for separating different components of primary cosmic-rays in the energy region above $\sim$ 10 TeV[32,33,34]. It has been shown that, while Hillas image parameters of atmospheric Cerenkov events recorded in an imaging Cerenkov telescope are able to separate protons from all other nuclei, the image parameters are rather inefficient in segregating nuclei of different masses. Our analysis confirms this fact even though we find that the image parameters like Width and Azwidth have the potential to separate the primaries into groups of light (protons and Ne) and heavy (Fe) nuclei. Several workers [34,35] have also investigated the efficiency of the temporal characteristics of atmospheric Cerenkov pulses for primary mass segregation. It has been reported that while some simple pulse shape parameters, like pulse rise time, full-width at half maximum and pulse decay time, carry information useful for cosmic-ray mass composition studies, the efficiency for primary mass segregation can not be quantified in view of the limited simulation database. A detailed feasibility study of cosmic-ray composition measurements with Cerenkov telescopes using fractal parameters of the recorded Cerenkov images has been presented earlier[23]. 

It has been shown that the mass sensitivity of all the parameters increases with the image size, possibly because of the decreasing event by event fluctuations with increasing primary energy($\sim$ size). However, a separation in more than two or three mass groups seems prohibited by the relatively poor mass resolution. Our results show that for primary hadrons of different masses with individual energies pre-selected to produce different amounts of Cerenkov light in the atmosphere, very efficient mass segregation is produced by the wavelet dimension $B_{6}$. As already noted, we find that the frequency distributions of $B_{6}$ values for the three primary nuclear species (p, Ne and Fe), considered by us in the simulations, peaks at distinctly separate values with negligible mutual overlap. To conclude, we have shown from our preliminary simulation-based study, that Cerenkov image shape parameters are able to separate primary cosmic-ray events into two or three groups of nuclei in terms of their mass but are inefficient in segregating the events in terms of primary mass on an event by event basis. However, we find that the wavelet dimension parameter $B_{6}$  of the Cerenkov image, has immense potential for event by event segregation(21,22,23) of the recorded events in terms of the primary mass. 

\section{Acknowledgement}
The author is grateful to Ramesh Koul, Head Astrophysical Sciences Division, BARC, Mumbai for his encouragement and overall support during the course of this work. Thanks are also due to Dr.A.K.Razdan for his help in developing the software used in this work and to Dr.K.K.Yadav for the help rendered during the course of data analysis.

\end{document}